# Optimal template banks

Bruce Allen[*]

*Max Planck Institute for Gravitational Physics (Albert Einstein Institute), Leibniz Universität Hannover,
Callinstrasse 38, D-30167 Hannover, Germany*

When searching for new gravitational-wave or electromagnetic sources, the $n$ signal parameters (masses, sky location, frequencies,...) are unknown. In practice, one hunts for signals at a discrete set of points in parameter space, with a computational cost that is proportional to the number of these points. If that is fixed, the question arises, where should the points be placed in parameter space? The current literature advocates selecting the set of points (called a "template bank") whose Wigner–Seitz (also called Voronoï) cells have the smallest covering radius ($\equiv$ smallest maximal mismatch). Mathematically, such a template bank is said to have "minimum thickness". Here, for realistic populations of signal sources, we compute the fraction of potential detections which are "lost" because the template bank is discrete. We show that at fixed computational cost, the minimum thickness template bank does *not* maximize the expected number of detections. Instead, the most detections are obtained for a bank which minimizes a particular functional of the mismatch. For closely spaced templates, the fraction of lost detections is proportional to a scale-invariant "quantizer constant" $G$, which measures the average squared distance from the nearest template, i.e., the average expected mismatch. This provides a straightforward way to characterize and compare the effectiveness of different template banks. The template bank which minimizes $G$ is mathematically called the "optimal quantizer", and maximizes the expected number of detections. We review optimal quantizer and minimum thickness template banks that are built as $n$-dimensional lattices, showing that even the best of these offer only a marginal advantage over template banks based on the humble cubic lattice.

DOI: 10.1103/PhysRevD.104.042005

## I. INTRODUCTION

Many searches for gravitational-wave and electromagnetic signals are carried out using matched filtering, which compares instrumental data to waveform templates [1–3]. Because the parameters of the sources are not known *a priori*, many templates are required, forming a grid in parameter space [4–7]. Like the mesh on a fishing net, the grid needs to be spaced finely enough that signals don't slip through. But if the grid has far more points than are needed, the computational cost becomes excessive. For this reason, a substantial technology has evolved to create these grids [8–14].

What choice of grid is optimal for a particular number of grid points? The literature on the topic answers the question as follows: select the grid which minimizes the largest distance between any point in parameter space and the nearest grid point [14].

If the grid is an $n$-dimensional lattice, then this choice corresponds to picking the lattice of minimum "thickness". That is to say, it selects the lattice whose Wigner-Seitz (WS) cells (also called Voronoï cells, Brillouin zones, and Dirichlet cells) have the smallest maximum radius. (An introduction to lattices and a description of the "classical" lattices may be found in Chapters 2 and 4 of [15].)

Here, we show that the minimum-thickness grid is not the best choice; it does not minimize the number of signals which are "lost" because of the discreteness of the grid. For that purpose, and provided that the grid points are not too widely separated, the best choice is the grid that minimizes the (normalized) second moment, which is the mean value of the squared distance (mismatch) to the nearest grid point. Mathematicians call such a grid the "optimal quantizer".

In this paper, we obtain simple expressions for the fraction of lost sources, in terms of the moments of the grid. If the grid is closely spaced, this expression only involves the second moment. For more widely spaced grids, we replace the usual quadratic approximation for the mismatch with a more accurate spherical ansatz [16]. The resulting expression for the fraction of sources lost involves all of the even moments, although in most cases the first

[*]bruce.allen@aei.mpg.de

Published by the American Physical Society under the terms of the Creative Commons Attribution 4.0 International license. Further distribution of this work must maintain attribution to the author(s) and the published article's title, journal citation, and DOI. Open access publication funded by the Max Planck Society.

half-dozen even moments are sufficient for an accurate approximation.

The main results of this paper apply to any grid of templates in parameter space. However, in many applications a regular grid is desirable; this may be systematically constructed as an $n$-dimensional lattice [15]. A lattice is obtained from a set of $n$ linearly-independent basis vectors, forming all linear combinations with integer coefficients. For a lattice, all WS cells are identical under lattice translation. In seven and nine dimensions, there are additional point sets which arise in this paper. These are tessellations whose WS cells have identical size and shape, but different orientations (via reflection and rotation).

The paper is structured as follows. In Sec. II we briefly review matched filtering, templates and template banks, the overlap between templates, and the mismatch function $m$ on parameter space. In Sec. III we show how, using the mismatch as a distance measure, the parameter space is broken up into WS cells surrounding each template. The radius of the smallest sphere which encloses one of these WS cells is called the covering radius (or WS radius) of the grid. In Sec. IV we review the conventional wisdom for template placement, which is to select the template grid points so as to minimize the covering radius $R$ for a given average WS cell volume. This minimizes a quantity known as the thickness of the lattice. Section V contains the main results of this paper: we calculate the fraction of detections which are lost because of the discreteness of the template bank. Minimizing the fraction of lost detections for a fixed number of templates is achieved by minimizing a particular functional of the mismatch given in Eq. (5.10). For closely spaced templates, this amounts to minimizing the second moment of the template grid as shown in Eq. (5.6). In the mathematical literature, grids which minimize this quantity are called "optimal quantizers". In Sec. VI we extend these results to the case where the putative signals are not uniformly distributed in parameter space. Lastly, in Sec. VII we discuss possible choices of template grids, and summarize the current state of knowledge about optimal quantizers when the grids are lattices or tessellations. This is followed by a short conclusion.

This paper mostly concentrates on the case of closely spaced templates (i.e., small mismatch). A companion paper [17] investigates the large-mismatch case in more detail, making use of the spherical ansatz to the mismatch [16].

Note that this paper uses "computing cost" as a surrogate for "number of templates" because for a simple one-pass search, the computing cost is proportional to the number of filter templates that are employed. There are also more complex multistage searches which have several different template banks, for example [18–23]. For those, the total computing cost depends upon many parameters, and is not directly proportional to "the number of templates". Nevertheless, the logic and results of this paper still apply to any particular stage of the search. For example, if other adjustable parameters are kept fixed, choosing an optimal quantizer for the template grid in any stage will minimize the number of lost signals in that stage.

## II. MATCHED FILTERING AND THE OVERLAP BETWEEN TEMPLATES

The classic signal detection problem is the following. We have instrumental or detector data in the form of a continuous or discretely sampled time series $s(t)$, which might or might not contain a weak signal with waveform $T(t)$ and unknown amplitude $\alpha$. The signal data stream is contaminated with zero-mean additive noise $n(t)$, so

$$s(t) = n(t) + \alpha T(t). \quad (2.1)$$

The problem is to identify (with some desired confidence) if the weak signal is present, and to estimate its amplitude.

The classic solution to this problem is called linear matched filtering [5,6,24–31]. This takes as input the data stream $s$ and the template $T$ and produces as output a single value, which is a positive-definite inner product

$$\rho = (Q_T, s), \quad (2.2)$$

where $Q_T = T/(T,T)^{1/2}$ is the "optimal filter" or "matched filter" associated with the template $T$. The inner product may be computed in the time or frequency domain; the optimal choice depends upon the properties of the noise. For example, if the noise is (at least second-order) stationary, then the inner product is

$$(A, B) = \int \frac{A^*(f)B(f)}{\mathcal{N}(f)} df, \quad (2.3)$$

where on the right-hand side the functions $A$, $B$ have been transformed to the frequency domain, and $\mathcal{N}(f)$ is the power spectrum of the noise $n(t)$. This inner product suppresses frequencies where the noise is large.

The expected value of $\rho$ (with a fixed signal and many instances of noise) is $\alpha(T,T)^{1/2}$, and the expected value of $\rho^2$ is $\alpha^2(T,T) + 1$. (A detailed calculation is given in Secs. 2 and 3 of [32].) The square of $\rho$ is called a "detection statistic"; large values indicate that the signal is likely present. Since the variance of $\rho$ is unity, the actual or expected values of $\rho$, $|\rho|$, and/or $\rho^2$ are called the signal-to-noise ratios (SNRs). For Gaussian noise, the statistical significance (log of the likelihood ratio) is proportional to $\rho^2$. This is reviewed in a signal-processing context in [33,34] and in the gravitational-wave (GW) context in [35,36].

If there was only a single possible signal waveform, then one template $T$ and one filter $Q_T$ would suffice. However, in most cases of interest, the signal waveform is dependent upon a number of unknown parameters. For example, the gravitational-wave signals produced by the inspiral of two masses depend upon the values of the masses, the sky location of the system, the spins of the two bodies, and the relative orientation and shape of the binary orbit. Here $n$ denotes the dimension of that parameter space and $\lambda^a$ with $a = 1, \ldots, n$ are coordinates on that space. We use $\lambda$ with no superscript to refer to the collection of these coordinate values.

Since the signal parameters are unknown, the template cannot match them precisely. This reduces the SNR compared to a perfect-match template, which has expected SNR $\rho_{\max} = \alpha(T, T)^{1/2}$. The mismatch $m$ is easy to characterize. In a mismatched template $T'$, with corresponding filter $Q_{T'}$, the expected (detected) SNR would be $\rho_{\det} = \alpha(T', T)/(T', T')^{1/2}$. The mismatch $m$ is the fractional loss

$$m = \frac{\rho_{\max}^2 - \rho_{\det}^2}{\rho_{\max}^2} = 1 - \frac{(T', T)^2}{(T, T)(T', T')}$$
$$= 1 - \cos^2\theta = \sin^2\theta, \quad (2.4)$$

where $\theta$ is the angle between $T$ and $T'$. It follows immediately that $m$ lies in the unit interval $m \in [0, 1]$.

In practice, many templates $T_i$ must be employed, where $i$ labels the template, corresponding to a signal with parameters $\lambda_i^a$. A set of $i = 1, \ldots, M$ discrete points $\lambda_i^a$ in parameter space is called a template bank with $M$ templates. Each template has an associated matched filter $Q_i = Q_{T_i}$. There are many techniques that can be used to construct such banks [8,12,13,37–39].

A search of the instrument data $s$ is carried out as follows. For each template in the bank, the SNR $\rho_i^2 = (Q_i, s)^2$ is computed. If any of the $\rho_i^2$ are above the detection threshold $\rho_D^2$, a detection is claimed [40]. In most cases, the parameters of the source are close to those of the template which registered the largest SNR.

While the signal parameters $\lambda^a$ might be close to the parameters $\lambda_i^a$ of one (or more) of the templates $T_i$, they will never be precisely the same; $\rho^2$ will be decreased by the parameter mismatch. If this causes $\rho^2$ to dip below the detection threshold in all templates, a potential detection would be missed. To quantify this, one uses Eq. (2.4) to define a mismatch function $m(\lambda) \geq 0$ everywhere on parameter space. Thus

$$m(\lambda) = \min_{i=1,\ldots,M}\left[1 - \frac{(T_\lambda, T_i)^2}{(T_\lambda, T_\lambda)(T_i, T_i)}\right], \quad (2.5)$$

where $T_\lambda$ is placed at $\lambda^a$ and $T_i = T(\lambda_i^a)$ ranges over all of the different templates in the bank.

Note: some papers on the topic of template banks use the symbol $m$ to denote the worst-case mismatch. Here, regardless of whether or not we indicate its dependence, the function $m = m(\lambda) = m(\lambda^1, \ldots, \lambda^n)$ denotes a function on parameter space. Its largest value (for a particular template bank) is denoted $m_{\text{worst}}$.

By definition, $m(\lambda)$ vanishes at the locations of the templates $\lambda = \lambda_i$. Thus, for $\lambda^a$ near the parameters $\lambda_i^a$ of the $i$th template, one has

$$m(\lambda) \approx r^2 = g_{ab}\Delta\lambda^a\Delta\lambda^b, \quad (2.6)$$

where $\Delta\lambda^a = \lambda^a - \lambda_i^a$, and $g_{ab}$ is a positive-definite symmetric quadratic form called the parameter-space metric. (We adopt the "Einstein summation convention": repeated indices are summed from 1 to $n$.)

This form of the mismatch $m(\lambda)$ is easily derived, assuming that the templates $T$ are smooth functions of the parameters $\lambda^a$, which implies that the mismatch is also a smooth function in the neighborhood of a template. Since $m(\lambda)$ is non-negative and vanishes at the template locations, those points are minima. Thus, expanding $m(\lambda)$ in a Taylor series about the template location, the constant and linear terms both vanish, so (generically) the quadratic term dominates.

In general, the parameter-space metric $g_{ab}(\lambda)$ will depend upon parameter-space position $\lambda$, but to simplify the treatment that follows, we will assume that the metric is constant (and thus describes a flat manifold). In practice, parameter space can usually be divided into regions where the metric is approximately constant. Via a linear transformation, one may then introduce new dimensionless coordinates on parameter space; with these coordinates the metric becomes the identity matrix, and $r^2$ is then the squared distance to the nearest template in standard Cartesian coordinates. In those coordinates, a lattice takes on its conventional appearance, for example as shown in Fig. 1. Our calculations and conclusions also apply to the image of the lattice in the original coordinates, however, it may have different appearance there. An explicit example of this construction may be found in Sec. 4 of [19].

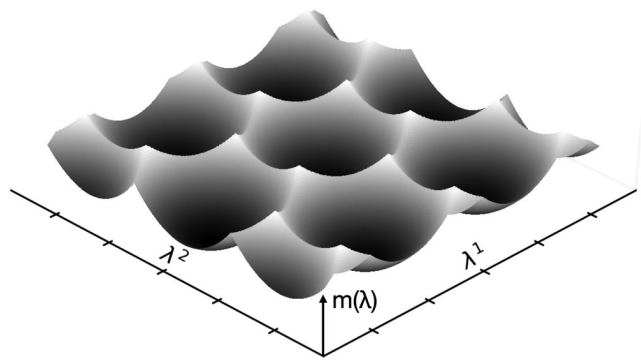

FIG. 1. The plane represents a 2-dimensional parameter space with coordinates $\lambda^1, \lambda^2$. Templates have been placed on an $A_2(\equiv A_2^*)$ lattice formed from the vertices of equilateral triangles. The vertical axis shows the mismatch $m(\lambda) = m(\lambda^1, \lambda^2)$, to the nearest template point. The zeros of $m$ are at the template locations. The discontinuities in the gradient of $m$ lie on the boundaries of the Wigner-Seitz (WS) cells, which are hexagons.

It follows from Eq. (2.5) that $r$ and $m$ are dimensionless quantities, so while we often refer to $r$ as a "distance", it is not a physical length. For this reason, in this paper, quantities which are independent of an overall rescaling of $r$ are called "scale invariant" rather than "dimensionless".

Most of the literature assumes the quadratic approximation for the mismatch given in Eq. (2.6). While this is valid provided that $m \ll 1$, it is unbounded above, whereas by definition the mismatch is bounded above by $m \leq 1$. Recent work [16] has shown that in many cases a better approximation to the mismatch is the "spherical ansatz"

$$m(\lambda) \approx \begin{cases} \sin^2 r & \text{for } r < \pi/2 \\ 1 & \text{for } r \geq \pi/2. \end{cases} \quad (2.7)$$

This approximation is also bounded to the correct range.

In what follows, we will investigate both approximations to the mismatch, and their consequences. For clarity, we will use $r^2 = g_{ab}\Delta\lambda^a\Delta\lambda^b$ to denote the quadratic approximation to the mismatch, and $\sin^2 r$ to denote the spherical approximation.

## III. WIGNER-SEITZ CELLS

Using the mismatch $m$ as a "distance measure", the parameter space for a given template bank may be partitioned into WS cells [41] which are in one-to-one correspondence with the templates: there is one cell surrounding each template. The WS cells are defined by the property that the points in a given cell have a smaller mismatch to that cell's template than to any other template.

These cells were also studied by Dirichlet [42], Voronoï [43–46] and Brillouin [47]. Since Voronoï was the first to investigate them in arbitrary dimension for arbitrary grids, it would be fair to use his name for them. However, in this paper we use "V" to denote volume, whereas "WS" is unambiguous.

Any given template bank has a maximum value of $r^2$, which here we denote by $R^2$ and call the "Wigner-Seitz radius". An example can be seen in Fig. 1. $R$ is also called the "covering radius"; it is the radius of the smallest ball [48] which encloses all points of the WS cell.

Traditionally, template banks have been constructed by (a) deciding how many templates are needed, then (b) placing the grid points so as to obtain the smallest possible value of $R$. This "traditional wisdom" corresponds to minimizing the thickness of the template grid.

## IV. THICKNESS, AND TRADITIONAL TEMPLATE PLACEMENT

The thickness (also called "covering density") $\Theta \geq 1$ is a scale-invariant quantity, defined as follows for an $n$-dimensional lattice. Let WS denote the Wigner-Seitz cell of the lattice point at the origin, and let $V(\text{WS})$ denote its $n$-volume. Let $R$ denote the WS "covering radius"; the maximum distance $r$ between the origin and any point in WS. Then the thickness is defined as

$$\Theta = \frac{V(B^n(R))}{V(\text{WS})}, \quad (4.1)$$

where $B^n(R)$ is an $n$-ball of radius $R$, and $V(B^n(R)) = \pi^{n/2}R^n/\Gamma(n/2+1)$ is its $n$-volume. Since the ball covers the WS cell, its volume cannot be smaller than that of the WS cell. Thus, by definition, $\Theta \geq 1$.

Note that the thickness is determined by the "shape" of the lattice and is independent of its scale. If all of the lattice spacings are rescaled by a factor $f$, then $R$ is rescaled by the same factor, and the volumes $V(B^n(R))$ and $V(\text{WS})$ are both rescaled by $f^n$, leaving $\Theta$ invariant.

The thickness is a measure of the way in which the balls of radius $R$ "overcover" space. If a ball of radius $R$ is placed around each lattice point, then $\Theta$ is the average number of balls in which a random point of parameter space lies, where we assume that the parameter space is large enough to contain many lattice points. A comprehensive review of classical lattices and their thickness and other properties may be found in Chapters 2 and 4 of [15].

The existing literature on template placement asserts that the best choice of grid is the one which has minimal thickness $\Theta$. The idea is that the available computing capacity determines the number $M$ of templates which can be employed, which means that $V(\text{WS})$ is fixed, since $M \cdot V(\text{WS})$ is the total volume of parameter space. Picking the grid with the smallest covering radius then ensures that the worst-case mismatch (in either the quadratic or the spherical approximation) is minimized. This in turn minimizes the thickness, Eq. (4.1). For example [14] states; "*The construction of optimal template banks for matched-filtering searches is an example of the sphere covering problem.... An optimal template bank therefore consists of the smallest possible number of templates that still guarantees that the worst-case mismatch does not exceed a given limit.*" A recent textbook [35], Sec. 7.2.1, says, "*The problem of constructing a grid in parameter space is equivalent to the so-called covering problem.... The optimal covering would have minimum possible thickness.*"

We will shortly show that, under reasonable assumptions, this choice, i.e., minimizing the thickness, is *not* optimal. If the goal is to maximize the expected number of detections for a given number of templates (i.e., at fixed computing power), it is better to place the templates to achieve the smallest average value of $r^2$ (if $r$ is small) or on some other combinations of moments (if $r^2$ is not small).

This fallacy, that the minimum thickness template bank is the best choice, may have a historical basis. The search for gravitational waves was the initial motivation for the study of matched-filter template banks in the 1990s, and for the development of the technology for template bank placement in the two decades that followed. Until the first signals were detected in late 2015 [49] the community focused on obtaining the most constraining "upper limits". These are upper bounds (with a stated statistical confidence) on the strength of different possible gravitational-wave sources; if a stronger source had been present, it would have been detected. For that purpose, obtaining the most constraining upper limits, which apply strictly over the entire parameter space, a minimum thickness template bank is optimal.

This has some important implications. Let us compare the cubic lattice $\mathbb{Z}^n$ to the $A_n^*$ lattice, which is an $n$-dimensional generalization of the 2-dimensional hexagonal lattice. The $A_n^*$ lattices are known to be the thinnest lattices for $n \leq 5$, and are either the thinnest or close to the thinnest classical lattices for dimensions $n \leq 17$, as discussed in [14]. For $6 \leq n \leq 24$, thinner lattices have been obtained, for example using a semidefinite programming approach (see Table 2 in [50]).

The thickness of $\mathbb{Z}^n$ is easily computed from Eq. (4.1), giving

$$\Theta[\mathbb{Z}^n] = \frac{(\pi n/4)^{n/2}}{\Gamma(\frac{n}{2}+1)}, \quad (4.2)$$

whereas the thickness of the $A_n^*$ lattice is [15,17]

$$\Theta[A_n^*] = \frac{(\pi n(n+2)/12)^{n/2}(n+1)^{(1-n)/2}}{\Gamma(\frac{n}{2}+1)}. \quad (4.3)$$

As the parameter-space dimension $n \to \infty$, the ratio asymptotes to $\Theta[\mathbb{Z}^n]/\Theta[A_n^*] \to 3^{n/2}/2\sqrt{n+1}$, and $\mathbb{Z}^n$ becomes much thicker than $A_n^*$. This is illustrated in Fig. 2 of [14]. For example, in eight dimensions, $\Theta[\mathbb{Z}^8] = 2\pi^4/3 \approx 64.94$, whereas $\Theta[A_8^*] = 20000\pi^4/531441 \approx 3.66585$. If lattice thickness were directly relevant, this would appear to give great advantage to the $A_8^*$ lattice. However, we will see shortly that when ranked by lost detections, these two lattices are quite similar.

## V. DETECTIONS LOST FROM TEMPLATE MISMATCH

We now examine how the choice of template locations influences the expected number of detections. Assume that the sources populate the parameter space uniformly (this can be relaxed, see Sec. VI) and have a distribution of expected SNR values (in perfectly matched templates) which is described by a population distribution function $P$, so that $dN = P(\rho^2)d\rho^2$ is the number of sources in the SNR range $(\rho^2, \rho^2 + d\rho^2)$. If the signal amplitude $\alpha$ is inversely proportional to distance $\ell$ (as is the case for GWs), then $\rho^2 \propto 1/\ell^2$. For sources uniformly distributed in a flat $d = 2$-dimensional Galactic disk, one therefore has $dN \propto \ell d\ell \propto \rho^{-4}d\rho^2$, and for sources uniformly distributed in flat $d = 3$-dimensional space, one has $dN \propto \ell^2 d\ell \propto \rho^{-5}d\rho^2$ [51]. Thus we take

$$dN = P(\rho^2)d\rho^2 = \frac{d}{2}N_D\left(\frac{\rho_D^2}{\rho^2}\right)^{d/2}\frac{d\rho^2}{\rho^2}, \quad (5.1)$$

where $d$ is the effective dimension [52] of the source distribution and $N_D$ is the total number of detectable sources (i.e., sources with SNR above the detection threshold $\rho_D$).

Suppose that the parameter space is densely covered with a very large number of closely spaced templates. In this case, the expected number of detections is

$$N_D = \int dN = \int_{\rho_D^2}^{\infty} P(\rho^2)d\rho^2. \quad (5.2)$$

This is the best-case scenario.

Now consider the more realistic case, where the templates have mismatch $m(\lambda)$. The expected number of detections is reduced, because some of the population, whose SNR would be only slightly above the threshold if there was a perfectly matching template, will fall below the detection threshold, due to mismatch losses to the nearest template. To be detectable the SNR must satisfy $\rho^2(1-m) \geq \rho_D^2$. So, the expected number of detections is

$$N_{\text{found}} = V^{-1}\int\int_{\rho_D^2/(1-m(\lambda))}^{\infty} P(\rho^2)d\rho^2 dV. \quad (5.3)$$

Here, $m(\lambda)$ denotes the mismatch to the nearest template, and the outer integral is over the $n$-dimensional parameter space. The volume element, in coordinates $\lambda$ for which the population distribution is uniform, is $dV = d\lambda^1 \cdots d\lambda^n = d^n\lambda$, and $V = \int dV$ is the $n$-volume of parameter space.

The number of detections "lost" because of the finite spacing of the template bank is the difference $N_{\text{lost}} = N_D - N_{\text{found}}$, which is therefore

$$N_{\text{lost}} = V^{-1}\int\int_{\rho_D^2}^{\rho_D^2/(1-m(\lambda))} P(\rho^2)d\rho^2 dV. \quad (5.4)$$

We now investigate several limits of this expression.

A simple limit is obtained if the template bank is closely spaced, so that everywhere in the parameter space the maximum mismatch $m$ is small compared to unity, implying that $m \approx r^2$. In this limit $\rho_D^2/(1-m) \approx \rho_D^2(1+r^2)$, so that

$$N_{\text{lost}} \approx V^{-1} \int \int_{\rho_D^2}^{\rho_D^2 + \rho_D^2 r^2(\lambda)} P(\rho^2) d\rho^2 dV$$

$$\approx V^{-1} \rho_D^2 P(\rho_D^2) \int r^2(\lambda) dV. \quad (5.5)$$

Making use of Eq. (5.1), the fraction of lost detections is then

$$\frac{N_{\text{lost}}}{N_D} \approx \frac{d}{2} V^{-1} \int r^2(\lambda) dV = \frac{d}{2} \langle r^2 \rangle, \quad (5.6)$$

where the final equality serves to define the "average second moment" $\langle r^2 \rangle$ of the template grid.

Thus, the number of "lost" detections is determined by the average value of the mismatch over the template bank. If the template bank is a lattice or tessellation in parameter space, then the fraction of lost detections (compared with a very closely spaced template bank) is

$$\frac{N_{\text{lost}}}{N_D} \approx \frac{d}{2} V_{\text{WS}}^{-1} \int_{\text{WS}} r^2 dV, \quad (5.7)$$

where the integral is over a single WS cell of volume $V_{\text{WS}}$, and we have assumed that the parameter space contains a large number of such cells.

The computational cost is determined by the number of templates at which the SNR is calculated, so fixing the computational cost is equivalent to fixing the number of templates, or fixing the volume $V_{\text{WS}} = V(\text{WS})$. Thus, *at fixed computational cost, if the templates are closely spaced, the number of lost signals is minimized by minimizing the average mismatch over the template bank*. In the mathematical literature [53], the quantity $\int_{\text{WS}} r^2 dV / V_{\text{WS}}$ is called the "normalized second moment of the lattice". *The lattice that minimizes this quantity in n dimensions, for fixed Wigner-Seitz cell volume $V_{\text{WS}}$, is called the optimal n-dimensional quantizer* [15].

To compare lattices, it is conventional to introduce the scale-invariant second moment of the lattice,

$$G = \frac{1}{n} \frac{\int_{\text{WS}} r^2 dV}{(\int_{\text{WS}} dV)^{1+\frac{2}{n}}}. \quad (5.8)$$

In the literature, $G$ is called "the mean squared error" or "dimensionless second moment" or "quantizer constant" of the lattice (but please note [53]). It is minimized by the optimal quantizer.

For a closely spaced template bank,

$$\frac{N_{\text{lost}}}{N_D} \approx \frac{nd}{2} (V_{\text{WS}})^{2/n} G[\text{lattice}]. \quad (5.9)$$

Thus, the relative number of lost signals at fixed computing cost for two closely spaced lattices can be estimated from the ratio of their quantizer constants $G$.

If the mismatch $m$ is not small, then the fraction of lost signals also depends upon the higher moments of the grid. One example is the search for continuous gravitational waves (CW) from rapidly spinning neutron stars in the Galactic disk, which have approximately a $d = 2$-dimensional distribution [52]. The parameter space for an uninformed search is very large, and so these searches are computationally limited, and often carried out at large mismatch. Let us assume a $d$-dimensional source distribution, and use Eq. (5.1) to evaluate the inner integral in Eq. (5.4). The fraction of lost detections is then

$$\frac{N_{\text{lost}}}{N_D} = V^{-1} \int (1 - (1 - m(\lambda))^{d/2}) dV$$

$$\approx V^{-1} \int (1 - \cos^d r) dV. \quad (5.10)$$

In the final line, we have used the spherical ansatz Eq. (2.7), assuming that the WS radius $R \leq \pi/2$.

If the grid is a lattice, then the integral can be replaced by the integral over a single WS cell. For WS radius $R \leq \pi/2$ we obtain

$$\frac{N_{\text{lost}}}{N_D} \approx \frac{d}{2} \langle r^2 \rangle - \frac{d(3d-2)}{24} \langle r^4 \rangle + \frac{d(15d^2 - 30d + 16)}{720} \langle r^6 \rangle$$

$$- \frac{d(105d^3 - 420d^2 + 588d - 272)}{40320} \langle r^8 \rangle$$

$$+ \cdots, \quad (5.11)$$

where

$$\langle r^p \rangle = V_{\text{WS}}^{-1} \int_{\text{WS}} r^p dV \quad (5.12)$$

denotes the normalized $p$th moment of the WS cell.

Several authors have investigated how continuous gravitational-wave searches should be structured, to provide maximum sensitivity at fixed computing cost [18,54–56]. Their results, which assume closely spaced templates, foreshadow ours. While those papers and later work do not explicitly discuss the optimization of a template bank or lattice, they indicate that the optimal search-parameter choices (for example, stack-slide time baseline) and achievable population-averaged sensitivity [18,57–59] are determined by the average value that the mismatch takes over the template bank, and not by the bank's thickness.

Note that our derivation assumes the actual detector SNR $\rho$ can be replaced by its expected value, or equivalently that the effects of detector noise may be neglected; the actual distribution of SNR in the detector would marginalize over the noise. For Gaussian noise and reasonable detection thresholds (say $\rho_D \approx 8$) this makes a negligible difference.

TABLE I. The smallest scale-invariant second moment (smallest $G$) lattices currently known for low dimension $n$. Also listed are the thinnest (smallest $\Theta$) classical lattices (but note that thinner nonclassical lattices have been constructed for $n \geq 6$, see text and [50]). Respectively, these minimize the second moment and covering radius at fixed WS-cell volume; smaller values are boldfaced. A minimum-$\Theta$ template bank yields the most constraining (strict) upper limit on source amplitude; a template bank of minimum $G$ yields the fewest "lost" detections (for small mismatch). In dimensions 7 and 9, the best known quantizers are the nonlattice tessellations $D_7^+$ and $D_9^+$, see text, footnotes, and [60,61] for details.

| Dimension $n$ | Lattice | Thickness $\Theta$ | Second moment $G$ | Ball limit on $G$[a] |
|---|---|---|---|---|
| 1 | $A_1^* = \mathbb{Z}$ | 1[b] | 0.083333[c] | 0.0833333 |
| 2 | $A_2^*$ | 1.2092[b] | 0.080188[c] | 0.0795775 |
| 3 | $A_3^*$ | 1.4635[b] | 0.078543[c] | 0.0769669 |
| 4 | $D_4$ | 2.4674[d] | **0.076603**[c] | 0.0750264 |
|   | $A_4^*$ | **1.7655**[b] | 0.077559[c] |  |
| 5 | $D_5^*$ | 2.4982[d] | **0.075625**[c] | 0.0735161 |
|   | $A_5^*$ | **2.1243**[b] | 0.076922[c] |  |
| 6 | $E_6^*$ | 2.65207[e] | **0.074244**[n] | 0.0723009 |
|   | $A_6^*$ | **2.5511**[b] | 0.076490[f] |  |
| 7 | $D_7^+$ | 4.7248[m] | **0.072734**[g] | 0.071298 |
|   | $E_7^*$ | 4.1872[h] | **0.073116**[o] |  |
|   | $A_7^*$ | **3.0596**[b] | 0.076187[i] |  |
| 8 | $E_8$ | 4.05871[d] | **0.071682**[c] | 0.0704536 |
|   | $A_8^*$ | **3.6658**[b] | 0.075971[i] |  |
| 9 | $D_9^+$ | 4.3331[m] | **0.071103**[g] | 0.069731 |
|   | $AE_9$ [j] | 10.3278[p] | **0.071622**[j] |  |
|   | $A_9^*$ | **4.3889**[b] | 0.075816[i] |  |
| 10 | $D_{10}^+$ | 7.7825[m] | **0.070813**[k] | 0.0691043 |
|   | $A_{10}^*$ | **5.2517**[b] | 0.075704[i] |  |
| 11 | $A_{11}^{3\ 1}$ | 27.089[l] | **0.070426**[l] | 0.0685548 |
|   | $A_{11}^*$ | **6.2813**[b] | 0.075624[i] |  |
| 12 | $K_{12}$ | 17.7834[d] | **0.070095**[l] | 0.0680682 |
|   | $A_{12}^*$ | **7.5101**[b] | 0.075568[i] |  |
| 13 | $D_{13}^*$ | 33.1285[q] | **0.074874**[r] | 0.0676338 |
|   | $A_{13}^*$ | **8.9761**[b] | 0.075531[i] |  |
| 14 | $D_{14}^*$ | 60.2442[q] | **0.074955**[r] | 0.0672433 |
|   | $A_{14}^*$ | **10.727**[b] | 0.075507[i] |  |
| 15 | $D_{15}^*$ | 66.0002[q] | **0.075040**[r] | 0.0668899 |
|   | $A_{15}^*$ | **12.817**[b] | 0.075495[i] |  |

[a]Eq. (7.1).
[b]Eq. (4.3) or Table 2.1 in Conway and Sloane [15].
[c]Table 2.3 in [15].
[d]Table 2.1 in [15].
[e]From $R$ and det following Ch. 4 Eq. (126) in [15].
[f]Ch. 21 Eq. (51) in [15].
[g]Nonlattice packing (tessellation). See Agrell and Eriksson [60] and Notes on Ch. 2 in [15]. Exact values were found by Dutour Sikirić [62] using the methods of [63] and are $G[D_7^+] = 178751/2457600$ and $G[D_9^+] = 924756607/13005619200$.
[h]From $R$ and det following Ch. 4 Eq. (115) in [15].
[i]Appendix of Allen and Shoom [17].
[j]Nonclassical lattice, discovered numerically by Agrell and Eriksson [60] [Eq. (31)] and denoted with their initials; exact solution in [61].
[k]Identified in [60]; exact value from Dutour Sikirić *et al.* [63].
[l]Exact value from [63].
[m]Text before Ch. 4 Eq. (94) of [15] with last paragraph [60] Sec. 3.
[n]Estimated for Table 2.3 of [15]; exact value from Worley [64].
[o]Estimated for Table 2.3 of [15]; exact value from Worley [65].
[p]E. Agrell, private communication. The deep holes of the lattice $B$ of Eq. (31) of [60] have the form $(\pm 1, 0, 0, 0, 0, 0, 0, 0, \pm a)$, where $a \approx 0.573$, giving a covering radius $R = (1 + a^2)^{1/2}$. The volume of the WS cell is $\det(B) = 2a$, giving $\Theta = \pi^{9/2} R^{9/2}/\det(B)\Gamma(11/2) = 16\pi^4 (1 + a^2)^{9/2}/945a$.
[q]$\Theta = V(B^n(R))/\text{Vol}(V(0))$ where $R$ and $\text{Vol}(V(0))$ are from Ch. 21 Eqs. (47) and (48) in [15].
[r]From Ch. 21 Eqs. (47) and (48) in [15].

## VI. NONUNIFORM POPULATION DENSITY AND THRESHOLD

Section V assumes that the sources populate the parameter space uniformly, and that the detection threshold is independent of the source type. Both of these assumptions can be dropped. Let $dN = P(\rho^2, \lambda) d\rho^2 dV$ be the expected number of sources in the parameter-space volume $dV$ with SNR in the range $(\rho^2, \rho^2 + d\rho^2)$. Then the more general result is

$$N_{\text{lost}} = V^{-1} \int \int_{\rho_D^2(\lambda)}^{\rho_D^2(\lambda)/(1-m(\lambda))} P(\rho^2, \lambda) d\rho^2 dV. \quad (6.1)$$

The small-mismatch limit is easily obtained, showing that the number of lost signals is determined by the average value of the mismatch, appropriately weighted by the number of sources in that region of parameter space.

Note that here and in the previous section, in order to write these expressions as integrals, we have assumed that the source probability distribution $P(\rho^2, \lambda)$, when considered as a function of $\lambda$, varies over scales much larger than the template spacing. This assumption is satisfied for the searches that we are familiar with, but we do not know if that will always be the case.

## VII. CHOICE OF OPTIMAL LATTICE

In a given parameter-space dimension $n$, what lattice is optimal? In contrast, how much is lost if a nonoptimal lattice is selected? How would this compare with a lattice selected for minimum thickness (smallest covering radius)?

If the mismatch is large (i.e., the quadratic approximation cannot be used), then these questions are not easily answered. In a companion paper [17] we have computed the fraction of lost detections for the cubic lattice $\mathbb{Z}^n$ and the $A_n^*$ lattice, which is an $n$-dimensional generalization of the hexagonal lattice.

If the mismatch is small enough that the quadratic approximation $m \approx r^2$ is valid, then we can see from Eq. (5.9) that at fixed computing cost (constant $V_{\text{WS}}$) the lattice which minimizes the number of lost signals is the optimal $n$-dimensional quantizer. By definition, this minimizes the scale-invariant second moment $G$ defined in Eq. (5.8).

To compute the relative fraction of sources lost by two lattices, we only need to compute the ratio of their $G$ values. Moreover, it is easy to compare any lattice with the cubic lattice. Thanks to the factor of $1/n$ which appears in the definition Eq. (5.8), the cubic lattice $\mathbb{Z}^n$ has a scale-invariant second moment $G = 1/12 = 0.08333\cdots$ which is independent of dimension $n$.

Table I summarizes the current state of knowledge for dimensions $n < 16$: it shows the lattices which have the smallest-known scale-invariant second moment $G$, along with references. In dimensions seven and nine, the best currently known quantizers are the nonlattice tessellations $D_7^+$ and $D_9^+$; for completeness we have also listed the best currently known lattice quantizers in those dimensions. For comparison, we have also listed the classical lattices with the smallest known thickness. (For $n \geq 6$, thinner lattices have been constructed numerically, by semidefinite optimization in the space of lattices; see Table 2 of [50]).

What is remarkable, and immediately visible from Table I, is that for small mismatch, where the quadratic approximation $m = r^2$ applies, the best lattices, with typical $G \approx 0.07$, have only a very marginal advantage in terms of lost signals when compared with the humble cubic lattice $\mathbb{Z}^n$, with $G = 1/12 \approx 0.083$.

The second moment of the ball $B^n$ provides a lower limit for the scale-invariant second moment $G$. One obtains

$$G[\text{Any grid}] \geq G[B^n] = \frac{\Gamma(n/2+1)^{2/n}}{\pi(n+2)}. \quad (7.1)$$

One can evaluate $G[B^n]$ in the $n \to \infty$ limit using Stirling's approximation, showing that $G > 1/2\pi e \approx 0.0586$. Note that [66] conjectures but does not prove a more constraining bound obtained by removing the part of the ball outside certain flat faces.

This means that in comparison with the cubic lattice $\mathbb{Z}^n$, for closely spaced templates the best choice of grid can *at most* reduce the fraction of lost signals by a factor of

$$\frac{N_{\text{lost}}[\mathbb{Z}^n]}{N_{\text{lost}}[\text{Best } n\text{-grid}]} \leq \frac{G[\mathbb{Z}^n]}{G[B^n]} < \frac{\pi e}{6} \approx 1.423. \quad (7.2)$$

In practice, the factor is substantially smaller than this. For example, in four dimensions the best known quantizer is $D_4$, which in comparison with the $\mathbb{Z}^4$ lattice would reduce the fraction of lost detections by about 8%, since $G[D_4]/G[\mathbb{Z}^4] \approx 0.919$. In eight dimensions the best known quantizer is $E_8$, whose fractional advantage over the $\mathbb{Z}^8$ lattice is about 14%, since $G[E_8]/G[\mathbb{Z}^8] \approx 0.860$.

For the $\mathbb{Z}^n$ lattice with closely spaced templates, the fraction of lost signals in Eq. (5.9) simplifies. Since $G = 1/12$, we have

$$\frac{N_{\text{lost}}}{N_{\text{D}}} \approx \frac{nd}{24}(V_{\text{WS}})^{2/n} = \frac{d}{6}R^2 = \frac{d}{6}m_{\text{worst}}, \quad (7.3)$$

where $R$ is the WS radius and $m_{\text{worst}}$ is the worst-case mismatch in the quadratic approximation. Thus, for a $d = 3$-dimensional distribution, at a worst-case mismatch of 20%, about 10% of signals would be lost.

## VIII. CONCLUSION

In this paper, we have shown in Eq. (5.10) how to quantify the fraction of detections which are lost because of the discreteness of a template bank; these sources could have been detected had the templates been more finely spaced. The fraction depends upon the properties of the source distribution and upon the placement of the templates. If the templates are not too far apart, the latter dependence is through the average value of the mismatch (second moment of the distance) as in Eq. (5.7).

For simplicity, our source models Eq. (5.2) assume time-independent source distributions with "Euclidean" volume measures. This is sufficient if sources are not at cosmological distances, so that the large-scale geometry of spacetime does not influence the measure, and if the sources are closer than $c\tau$, where $\tau$ is the time scale on which the properties of the source distribution evolve, and $c$ is the speed of light. Future generations of gravitational-wave detectors will have a reach which extends to the Hubble radius, and will study sources which have significant evolution over redshifts $z \lesssim 10$. For those, a precise estimate of "lost" sources may require population models that incorporate source and/or cosmological evolution.

To maximize the expected number of signal detections for a given number of templates, we have shown that a template bank must minimize the average of a function of the mismatch $m$, given in Eq. (5.10). For closely spaced templates, where the mismatch reduces to the squared distance to the nearest template, this corresponds to choosing a grid which is the "optimal quantizer" as in Eq. (5.9), which minimizes the average value of the squared distance to the closest point in the template bank. This contrasts with standard wisdom, which holds that the optimal choice of template bank is the one which minimizes the covering radius (or equivalently, the thickness).

Template bank thickness is relevant for upper limits, but it is necessary to distinguish between two types of upper limits: "strict" and "population-averaged". Strict upper limits apply at every point in parameter space, whereas population-averaged upper limits only apply on average (with the stated confidence) to the entire population. The literature contains examples of both. Sometimes (for example see [67–70]) both variants are given in the same paper. While the thinnest template bank will give the most constraining strict upper limit, it does not maximize the expected number of detections, and is probably also not optimal for the population-averaged upper limits.

Often, template banks are constructed as regular lattices. To compare two lattices in the closely spaced case, and to identify which choice maximizes the expected number of detections for a fixed number of templates, one need only compare the scale-invariant second moments (quantizer constants) $G$ of the lattices. The ratio of $G$ for two lattices is proportional to the relative numbers of "lost" detections at fixed computing cost, as can be seen from Eq. (5.6).

This has an important consequence for the humble cubic lattice $\mathbb{Z}^n$. While it has much thinner and more sophisticated cousins such as $A_n^*$, the ratio of their quantizer constants $G$ is not far from unity. This can be seen from Table I, keeping in mind that for the cubic lattice, $G = 1/12 \approx 0.08333$ in any number of dimensions. Hence, in many cases, the small marginal returns do not justify the additional complication that noncubic lattices entail.

It is easy to visualize why the cubic lattice $\mathbb{Z}^n$ is so thick; the corners of the cube "stick out", giving it a large covering radius. This makes it a poor choice for obtaining strict upper limits, because a signal hidden in one of those distant corners at radius $R$ could have much larger amplitude than the bulk of the population, yet might still go undetected. However, if the goal is detection (or a population-averaged upper limit), this does not matter. The volume in the corners is quite small, which in turn means that the expected number of signals lost there is also small [71].

The use of an optimal quantizer template bank has a further advantage; to lowest order it eliminates parameter-dependent bias in the number of detections. To see why, recall that WS cells are polytopes; convex regions bounded by planes. Being nonspherical, these "break the symmetry" between different waveform parameters. For example, a WS cell with large thickness could potentially bias a search, reducing the number of detections more than the average for signals whose parameters are shifted in the direction of the WS cell vertex most distant from the origin. Remarkably, there is a beautiful theorem by Zamir and Feder [72], that the optimal quantizer has a correlation matrix $C^{ab}$ proportional to the identity. This means that the optimal quantizer lattice does not single out any preferred directions in parameter space; all vectors are eigenvectors of $C^{ab}$, with the same eigenvalue. An explicit example may be found in [61], where the second-moment tensor $U^{ab}$, which is proportional to the correlation matrix, is calculated in closed form. Thus, employing an optimal quantizer template bank might ease astrophysical interpretation, by reducing the detection bias associated with inequivalent directions in parameter space. (Note that the cubic lattice shares this property, since its correlation matrix is also proportional to the identity.)

There are many types of computationally limited signal searches for which these results are relevant. For example, it is currently not possible to do an all-sky search for gamma-ray pulsations in binary systems, or for continuous gravitational waves from neutron stars in binary systems. The parameter space here (counting dimensions in parentheses) includes sky position (2) and frequency and spindown (2). For circular orbits one additionally has orbital period, inclination angle, and modulation depth (3); if the orbit is eccentric, then two additional parameters are needed. So in this case, the parameter space is 7- or 9-dimensional [73]. The situation is even worse for gravitational-wave searches from binary inspiral systems where spin effects are significant; for circular orbit systems there are 14 parameters [6]. Current technology does not have the computational power to explore such large dimensional spaces, but advances in quantum computing may permit such searches in the future.

Throughout this paper, we have neglected effects that arise at the boundary or boundaries of the parameter space. While our optimization criterion can still be applied in the presence of boundaries, the use of a lattice template bank assumes that the parameter space volume is large enough that boundary effects may be neglected. In practice, as the dimension of the parameter space increases, boundary effects become more significant. If boundary effects can not be neglected, it may still be possible to employ lattices and the lattice approximations we have used, by dividing parameter space into regions with different effective dimension.

A companion publication [17] looks in more detail at the case where the templates are not closely spaced, making use of the spherical ansatz Eq. (2.7) to model the mismatch at large separation.

### ACKNOWLEDGMENTS

I am grateful to many colleagues who have discussed this topic with me over the years, including Ben Owen, Maria Alessandra Papa, Reinhard Prix, B. S. Sathyaprakash, and Andrey Shoom. I thank Erik Agrell for assistance with Table I in seven and nine dimensions and for detailed comments on the manuscript, and Mathieu Dutour Sikirić, who computed the exact values of $G[D_7^+]$ and $G[D_9^+]$ given in the footnote to Table I, and brought the thinnest known lattices of [50] to my attention. Finally, I thank the anonymous referee, whose insightful comments led to several significant clarifications and improvements to the manuscript.